\documentstyle[twocolumn,aps]{revtex}
 \input epsf
\begin{document}
%\draft
\twocolumn[\hsize\textwidth\columnwidth\hsize\csname@twocolumnfalse\endcsname

\title{Empirical properties of the variety of a financial
portfolio and the single-index model}
\author{Fabrizio Lillo and Rosario N. Mantegna} 
\address{Max-Planck Institut f\"ur Physik komplexer Systeme,
N\"othnizer Str. 38, D-01187 Dresden, Germany 
  and Istituto Nazionale per la Fisica della Materia, 
  Unit\`a di Palermo, Viale delle Scienze, I-90128 Palermo, Italy }

%\date{\today} 
\maketitle %\receipt{}

%----------------------------------------------------------------------
\begin{abstract} 
We investigate the variety
of a portfolio of stocks in normal and extreme days of market 
activity. We show that the variety carries information
about the market activity which is not present in the 
single-index model and 
we observe that the variety time evolution
is not time reversal around the crash days. 
We obtain the theoretical relation between the square
variety and the mean return of the ensemble return distribution
predicted by the single-index model. The single-index
model is able to mimic the average behavior of the square variety 
but fails in describing quantitatively the relation between the
square variety and the mean return of the ensemble distribution.
The difference between empirical data and 
theoretical description is more pronounced for large positive values
of the mean return of the ensemble distribution. Other significant deviations
are also observed for extreme negative values of the mean return.
\end{abstract}
%----------------------------------------------------------------------
\vskip2pc]

%\newpage
\narrowtext
\section{INTRODUCTION}

Financial markets can be regarded as model complex systems
\cite{Anderson88}. They are 
open systems composed of many non-equivalent sub-units 
interacting in a nonlinear way. They are continuously monitored 
and a huge amount of carefully recorded financial data 
are now accessible for analysis and modeling of market
micro-structure. This allows to perform empirical analyses
elucidating statistical regularities that can be used to 
test models of financial activities \cite{Proc,MS00,BP00}. These tests 
provide information about the strengths and weaknesses of the
various models pointing out the aspects that need to be 
improved to obtain a better model.

Stylized facts observed in financial markets mainly   
refer to the statistical properties of asset returns
and volatility and to the degree and nature of 
cross-correlation between different assets traded 
synchronously or quasi synchronously and belonging to
given portfolios. Recently, we have proposed to
model the different behavior 
observed in the stock returns of a portfolio by considering 
the statistical properties (shape, moments, etc.) of the 
ensemble return distribution of stocks simultaneously traded
in a market. Our studies \cite{Lillo99,Lillo00a,Lillo00b} 
have shown that the statistical properties
of the ensemble return distribution are roughly conserved 
in normal days of activity of the market whereas during crash
and rally they change in a systematic way. 

The single-index model \cite{Elton,Campbell} is not adequate to model some of
these findings. Specifically, it fails in describing the
statistical properties of the standard deviation (called by us 
variety) of the 
ensemble return distribution and it misses to 
quantitatively reproduce the symmetry breaking 
of the empirical return distributions observed during crash 
and rally days \cite{Lillo99}. On the other hand the same model is a rather
attractive because it describes pretty well several
stylized facts related to the first moment 
of the ensemble return distribution.

In the present study we investigate the empirical behavior
of the variety with respect to the theoretical
predictions of the single-index model. We find that variety
is only mimicked at a ``zero-order" by the single-index model
and significant discrepancies are observed in the statistical properties
of this variable both in extreme days and in periods of normal activity 
of the market.

\section{The ensemble return distribution in normal and 
extreme days}

In our empirical analysis we investigate the statistical properties
of the ensemble return distribution obtained for a portfolio of stocks
traded in a financial market.
The investigated market is the New York Stock Exchange (NYSE) during the 
12-year period from January 1987 to December 1998. This time period
comprises 3032 trading days. Here we present empirical analyses 
of two different sets of stocks. The first is the set of all the stocks
traded in the NYSE. For this statistical ensemble the number of stocks
is not fixed because the total number of assets $n$ traded in the NYSE is 
rapidly increasing in the investigated time period and ranges from 
$1128$ in 1987 to $2788$ in 1998. The second set is the set of $1071$
stocks which are continuously traded in the NYSE in the considered 
period. The total number of financial records processed exceeds 
$6$ millions. 

The variable investigated in our analysis is the daily price return, 
which is defined as
\begin{equation}
R_i(t)\equiv\frac{Y_i(t)-Y_i(t-1)}{Y_i(t-1)},
\end{equation}  
where $Y_i(t)$ is the closure price of $i-$th asset at day $t$ ($t=1,2,..$). 
In our study we 
consider only the trading days and we remove the weekends and the holidays 
from the data set. Moreover we do not consider price returns which are 
in absolute values greater than $50\%$ because some of these returns might 
be attributed to errors in the database and may affect in a considerable 
way the statistical analyses. For each set of stocks, we extract the $n$ 
returns of the $n$ stocks at each trading day and we consider the  
probability density function (pdf) of price returns. 
The distribution of these returns describes the 
general activity of the market at the selected 
trading day. In the periods of normal activity of the market, 
the central part of the distribution is conserved for a long 
time. In these periods the shape of the distribution 
is systematically non-Gaussian and approximately symmetrical \cite{Lillo00b}.
During crashes and rallies the ensemble return distribution 
changes abruptly shape. In a previous 
study \cite{Lillo00a} we have shown that the 
ensemble return distribution becomes asymmetric in critical days.
Specifically, in crash days the ensemble return distribution
becomes negatively skewed whereas in rally days the distribution becomes
positively skewed. The change of the symmetry properties is not 
the only change of the pdf observed in crash and rally days. In fact 
during critical days the central moments of the pdf assume
values rather different from the typical ones.  

To illustrate the change of the distribution in crash and rally days
and in the nearby time periods
we select the three biggest crashes present during the time period 
of our database. They correspond to -- (i) the black
Monday crash of  19th October 1987 when the Standard and Poor's 500 
index had a $-20.4\%$ return, (ii) the crash of 27th October 1997 when the 
Standard and Poor's 500 index had a $-6.9\%$ return, and (iii) 
the crash occurring at 31st August 1998 when the 
Standard and Poor's 500 index had a $-6.8\%$ return. 
Related to these crash days there are also relevant rally days.
This is because the days of greatest 
rallies of our
database occur just one or few days after crashes. In the 1987 time period,
in addition to the rally days, a second crash of $-8.3\%$
of the Standard and Poor's 500 index occurred at 26th October 1987 .  
The statistical behavior of stock market indices during crashes 
has also been investigated under a different perspective in 
Ref. \cite{Johansen98}.
 
Figure 1 shows the contour plot of the logarithm of ensemble 
return distribution of the three above mentioned
crash days in a $200$ trading days time interval centered at 
day of the crisis. The contour plots of the three crises show 
analogies and differences. 
An analogy is observed by considering that the time period after the 
crisis is clearly characterized by a degree of high instability in
the shape of the ensemble return distribution.
This is shown by the behavior of the contour lines which are more parallel
before crises (negative values of the trading day index in Fig. 1)
than after crises. An ``aftershock" period lasting more than 50 trading 
days is clearly detected in two of the three analyzed cases, specifically 
in the 1987 and in the 1998 crises. Another aspect of this
analogy is that the shape of the distribution tends to fluctuate  
significantly during the days immediately after the crises.  
A difference can be noted by considering that the onset of the crises
is almost abrupt for the 1987 and for the 1997
crises whereas a progressive modification of the shape of the ensemble return
distribution is detected in the time period before 
the 1998 crisis.
 
In order to characterize more quantitatively  the ensemble return
distribution at day $t$, we extract the  first two central moments
at each of the $3032$  trading days. Specifically, we consider the
mean and the standard  deviation of the ensemble return 
distribution defined as
\begin{eqnarray}
&&\mu (t)=\frac{1}{n}\sum_{i=1}^{n} R_i(t), \\
&&\sigma (t)= \sqrt{\frac{1}{n}\left(\sum_{i=1}^{n}
(R_i(t)-\mu(t))^2\right)}.
\end{eqnarray}

The mean value of price returns $\mu(t)$ quantifies  the general trend
of  the market at day $t$.  The standard deviation $\sigma(t)$,
i.e. the {\it variety} \cite{Lillo99,Lillo00b}
of the ensemble return distribution,
measures its width. 
A large value of $\sigma(t)$
indicates that different companies are characterized by rather
different returns at day $t$.  In fact in days of high variety
some companies perform great gains whereas others have great
losses. The variety of price returns is not constant and fluctuates in time.

Figure 2 shows the variety as a function of the trading day index for the
same crises shown in Fig. 1. The behavior qualitatively observed in Fig. 1
is now quantitatively shown in Fig. 2. The abrupt onset of the 1987 and 1997
crises is rather clear, whereas a progressive increase of the variety
is observed before the 1998 crisis. The presence of an aftershock 
period longer than 50 trading days is observed in all three cases. Wild 
fluctuations of the variety are observed immediately after 
the 1987 and 1998 crises. In summary during a crisis and in a 
long period after the crisis the variety of a portfolio of
stocks increases significantly. It is worth pointing out that the highest
value of the variety is {\it not} observed at the crash day but rather
at the day immediately after in all the considered cases.

The variety of a portfolio of stocks is not a variable that
is invariant to time reversal. Indeed, our analysis of these case 
studies suggests that the behavior of the market just before and 
just after crises is rather different with respect to the variety of the 
portfolio of stocks.

\section{Single-index model}

We now investigate the theoretical properties of the variety 
of a portfolio of stocks described in terms of a 
single-index model. The theoretical predictions will be compared with
the results of our empirical observations and with surrogate data
in Section 3.2.

The single-index model \cite{Elton,Campbell} is a
basic model of price dynamics in financial markets. It assumes that the
returns of all stocks are controlled by one factor,  usually called the
``market". In this model, for any stock $i$ we have  
\begin{equation}
R_i(t)=\alpha_i+\beta_i R_M(t)+\epsilon_i(t),  
\end{equation}  
where $R_i(t)$ and $R_M(t)$ are the return of the stock $i$ and of the 
``market" at day $t$ respectively, $\alpha_i$ and $\beta_i$ are two
real  parameters and $\epsilon_i(t)$ is a zero mean noise term
characterized by a variance equal to $\sigma^2_{\epsilon_i}$. The noise
terms of different  stocks are assumed to be uncorrelated
for $i\neq j$.  Moreover the
covariance between $R_M(t)$ and $\epsilon_i(t)$ is set to zero for any
$i$. In this model each stock is correlated with the market and the 
presence of such a correlation induces a correlation between any pair of stocks.

There are several possible choices concerning the statistical 
properties of the noise terms $\epsilon_i$. The customary choice 
is a Gaussian statistics 
\cite{Elton,Campbell} 
but also non-Gaussian statistics \cite{Bouchaud00} has been considered.
For the moment we do not specify the statistical properties 
of the noise terms explicitly and we only assume that the variance of each  
$\epsilon_i(t)$ is finite.

\subsection{Central moments in the single-index model}

In our study we perform both ensemble and time averages. In our
notation $<...>$ indicates ensemble averaged quantity, whereas
$[...]$ indicates time averaged quantity.

The mean of the ensemble return distribution $\mu(t) \equiv <R_i (t)>$ 
is given by
\begin{equation}
\mu(t)=<\alpha_i>+<\beta_i>R_M(t)+<\epsilon_i(t)>.
\end{equation}
The quantity $<\epsilon_i(t)>$ is proportional to the sum of $n$
uncorrelated random variables with zero mean and finite variance. 
The central limit theorem ensures that $<\epsilon_i(t)>$ is Gaussian 
distributed with zero mean and variance given by 
$<\sigma^2_{\epsilon_i}>/n$. The time average of the random 
variable $\mu(t)$ is given by
\begin{equation}
[\mu(t)]= <\alpha_i>+<\beta_i>[R_M(t)].  
\end{equation}
The determination of higher moments requires the 
calculation of time and ensemble variances of random
variables. In the following, for a random
variable $x_i$ we indicate its ensemble variance as 
$<\Delta x_i^2>\equiv <(x_i-<x_i>)^2>$ whereas for a 
random variable $x(t)$ we indicate its time variance as 
$[\Delta x(t)^2]\equiv [(x(t)-[x(t)])^2]$.

The time variance of $\mu(t)$ is
\begin{equation}
[\Delta\mu(t)^2]= <\beta_i>^2[\Delta R_M(t)^2]+
\frac{<\sigma^2_{\epsilon_i}>}{n}.  
\end{equation}
In deriving this equation we use the property that the
covariance between $R_M(t)$ and $\epsilon_i(t)$ is set 
to zero for any $i$. In a similar way we compute $<R^2_i(t)>$. 
By assuming that 
the ensemble covariance between two of $\alpha_i$,
$\beta_i$ and $\epsilon_i$ is zero, we find that 
the square of the variety $\sigma(t)$ of Eq. (3) is equal to 
\begin{eqnarray}
&&\sigma^2(t)=<\Delta \alpha_i^2>\\
&&+<\Delta \beta_i^2> R^2_M(t)+<\epsilon^2_i(t)>-
<\epsilon_i(t)>^2 \nonumber
\end{eqnarray}
Using Eq.(5) to express $R^2_M(t)$ as a function of $\mu(t)$ 
we rewrite Eq.(8) as
\begin{eqnarray}
&&\sigma^2(t)=<\Delta \alpha_i^2>+\frac{<\Delta \beta_i^2>}{<\beta_i>^2}(<\alpha_i>^2+
<\epsilon_i(t)>^2) \nonumber \\
&&+<\epsilon^2_i(t)>-<\epsilon_i(t)>^2 \nonumber \\
&&-2<\alpha_i>\frac{<\Delta
\beta_i^2>}{<\beta_i>^2}\mu(t)+\frac{<\Delta\beta_i^2>}{<\beta_i>^2}\mu^2(t)
\end{eqnarray}
The relation between the square variety and the mean ensemble return 
is therefore quadratic. This implies that the single-index model 
predicts a linear increase
of the variety for large absolute value of the mean $\mu(t)$
\begin{equation}
\sigma(t)\approx\frac{\sqrt{<\Delta\beta_i^2>}}{<\beta_i>}|\mu(t)| 
\end{equation}
The proportionality factor of Eq. (10) gives a measure of the 
inhomogeneity of the portfolio with respect to the market factor. 
Hence the increase of the 
variety for large values of $|\mu(t)|$ is due to the inhomogeneity
of the portfolio in following the market behavior. This result 
is independent of the statistics of the noise terms. 

Equation (10) is valid only for large values of $|\mu(t)|$. 
To obtain a general expression of the square variety as a function
of the single-index model parameters we need to make explicit 
the terms $<\epsilon_i^2(t)>$ and $<\epsilon_i(t)>^2$ of 
Eqs (8) and (9).
The
term $<\epsilon^2_i(t)>$ is proportional to the sum of the 
squares of $n$ independent random variables each with mean equals to 
$\sigma^2_{\epsilon_i}$ and variance which is dependent on the 
statistical properties of $\epsilon_i$. We indicate the time variance 
of $\epsilon_i^2(t)$ with $v_i$.
By applying the central limit theorem one can show that
$[<\epsilon^2_i(t)>]$ is equals to $<\sigma^2_{\epsilon_i}>$ 
and the variance of $<\epsilon^2_i(t)>$ is $<v_i>/n$. 

Let us now assume that the noise variables $\epsilon_i(t)$ are Gaussian. 
In this case the variance of $\epsilon_i^2$ is given by 
$v_i=2\sigma^4_{\epsilon_i}$. By using this property, we 
conclude that the first two central moments of the random variable 
$<\epsilon^2_i(t)>$ are for Gaussian noise terms
\begin{eqnarray}
[<\epsilon^2_i(t)>]=<\sigma^2_{\epsilon_i}>  \\
\left[\Delta<\epsilon^2_i(t)>^2\right]=\frac{2}{n}<\sigma^4_{\epsilon_i}>
\end{eqnarray}
The term $<\epsilon_i(t)>^2$ is the square of a single 
Gaussian variable and is distributed according to the Gamma function 
$f_{a,\nu}$ \cite{Feller} 
with $a=n/(2<\sigma^2_{\epsilon_i}>)$ and $\nu=1/2$. 
The mean of this term is $<\sigma^2_{\epsilon_i}>/n$ and 
the variance is $2<\sigma^2_{\epsilon_i}>^2/n^2$.
Hence, we conclude that 
\begin{equation}
<\epsilon^2_i(t)>-<\epsilon_i(t)>^2\approx <\epsilon^2_i(t)>.
\end{equation}
By using Eqs (8), (11-12) and (13) we can  explicitly write down the first 
two temporal moments of the random variable 
$\sigma^2(t)$ 

\begin{eqnarray}
&&[\sigma^2(t)]=<\Delta \alpha_i^2>+<\Delta \beta_i^2> [R^2_M(t)]+
<\sigma^2_{\epsilon_i}>\\
&&\left[\Delta(\sigma^2(t))^2\right]= \nonumber \\
&&<\Delta\beta_i^2>^2([R_M^4(t)]-[R_M^2(t)]^2) 
+\frac{2}{n} <\sigma^4_{\epsilon_i}>.
\end{eqnarray}
The validity of Eq. (14) is independent of the Gaussian 
assumption for the statistical properties of noise terms. On the other
hand, Eq. (15) is valid only for Gaussian noise
terms and for a market factor with finite fourth moment.

\subsection{Comparison with empirical and surrogate data}

In order to verify the relation between mean and variance predicted by the
single-index model we generate surrogate data of 
an ``artificial" market according to Eq. (4). The investigation of empirical
data and the associated study of surrogate data are performed by
considering the set of $1071$ stocks traded continuously in the NYSE 
in the time period $1987-1998$. By using the ordinary least square 
method we estimate the 
model parameters $\alpha_i$, $\beta_i$ and $\sigma^2_{\epsilon_i}$ 
for all the stocks of our ensemble. We recall that the best 
estimate of the model parameters does not depend on the statistical
properties of noise terms. In Table 1 we show the 
ensemble mean and standard deviation  of these parameters.
The Standard and Poor's 500 index is chosen by us as the 
market factor $R_M(t)$. It assumes the following values for 
the first two temporal moments
$[R_M(t)]=5.80~10^{-4}$ and $[\Delta R_M(t)^2]=1.02~10^{-2}$
in the investigated time period.
The time series of surrogate data are generated by using the 
above cited parameters and market factor.
To check the role of the statistical properties of the noise terms 
we consider two different choices for  
$\epsilon_i(t)$ --(i) noise terms with Gaussian statistics and (ii) 
noise terms with non-Gaussian statistics. For the case (ii) we
assume that $\epsilon_i(t)=\sigma_{\epsilon_i}w(t)$, where $w(t)$ 
is a random variable distributed according to a Student's {\it t} 
probability density function 
\begin{equation}
P(w)=\frac{C_\kappa}{(1+w^2/\kappa)^{(\kappa+1)/2}},
\end{equation}
where $C_\kappa$ is a normalization constant. Empirical investigations 
of real data 
\cite{Lux96,Gopikrishnan98,Gopi99} indicates a value 
between $4$ and $6$ for 
the power-law exponent of $P(w)$ for large values of $|w|$. 
In our simulation
we take the most leptokurtic distribution within this interval 
which corresponds to $\kappa=3$.

We analyze the surrogate data in the same way used to investigate
empirical data. In Fig. 3 we show the time series of the variety 
corresponding to the same time periods of Fig. 2 obtained for the
surrogate data with Student's $t$ noise terms. The time series
of Fig. 3 are rather different from the ones shown in Fig. 2. A similar
behavior is observed for the surrogate data with Gaussian noise terms.
In Fig. 3 the increases of the variety occurring at the crash day 
is still evident in all cases but surrogate
data are not able to describe
the long aftershock period observed in empirical data. In other words
the temporal asymmetry with respect to crash day observed in 
empirical data is not reproduced by the single-index model
in spite of the fact that the behavior of $\mu(t)$ is well
reproduced by the single-index model for all the considered time periods.
This suggests that the variety of a stock portfolio is more
sensitive to temporal asymmetry than the market factor $R_M(t)$.
Moreover, a detailed analysis of  Fig. 2 shows that 
the day of highest variety is always the
crash days in the surrogate data whereas in Section 2 we noted that
the day of highest variety is the day after the crash in empirical 
data.
 
To make a quantitative comparison between empirical data, 
surrogate data and the theoretical predictions of the 
single-index model we 
study some statistical properties of $\mu(t)$  and $\sigma^2(t)$.
Table 2 shows the temporal mean and the standard deviation of $\mu(t)$ 
for (i) empirical data, (ii) single-index theoretical 
prediction based on Eqs (6) and (7) and using the model parameters of
Table 1,
(iii) surrogate data with Gaussian statistics of the noise terms, and (iv)
surrogate data with Student's {\it t} statistics of the noise terms.
The agreement between the results obtained for empirical and 
surrogate data with the theoretical predictions of the single-index
model is pretty good. As stated in section 3.1 the mean and the
standard deviation of $\mu(t)$ are not significantly affected by
the statistical properties of the noise terms. This result has 
been observed in Ref. \cite{Lillo00b} in numerical simulations.

In Table 3 we show the values of the mean and standard 
deviation of $\sigma^2(t)$ for the same sets of data and 
theoretical predictions as in Table 2. It is worth pointing out 
that the theoretical value of the standard deviation of $\sigma^2(t)$ 
is obtained under the assumption of Gaussian statistics for 
noise terms. The same quantity diverges under the assumption
of Student's-$t$ noise terms with $\kappa \le 4$. We find that
the mean value of $\sigma^2(t)$ is approximately reproduced 
by both Gaussian and Student's {\it t} surrogate data.
The theoretical estimation of Eq. (14) with the
model parameters of Table 1 is also close to the 
empirical value. However a different conclusion is 
obtained for the standard deviation of
$\sigma^2(t)$. None of the values obtained from the theoretical
estimation and numerical simulation of surrogate data 
is able to explain the value observed in empirical data.
This is not due to the fact that the theoretical prediction is
inaccurate because the theoretical prediction well describes 
the case of surrogate data with Gaussian noise terms. Hence, the 
different value detected in the empirical analysis of the 
standard deviation of $\sigma^2(t)$ is a clear manifestation
of the limit of the single-index model. This model is able to describe the
behavior of $\mu(t)$ but it is not able to describe the square 
variety of the portfolio properly.

To support the above conclusion we now consider in more detail
the relation between $\sigma^2(t)$ and $\mu(t)$
of the ensemble return distribution. The values of the 
model parameters are such that
Eq. (9) can be approximated as
\begin{eqnarray}
&&\sigma^2(t) \approx <\sigma^2_{\epsilon_i}>-2<\alpha_i>\frac{<\Delta
\beta_i^2>}{<\beta_i>^2}\mu(t) \nonumber \\
&&+\frac{<\Delta\beta_i^2>}{<\beta_i>^2}\mu^2(t).
\end{eqnarray}
Figure 4 shows the square of the variety $\sigma^2(t)$ as a function of the
mean $\mu(t)$ for each trading day of the investigated period. In the figure
each circle refers to a different trading day. The filled circle are 
obtained from the real data whereas the empty circles 
are obtained from surrogate data generated according to Eq. (4) and 
by using a Gaussian statistics for the noise terms. The dashed line is obtained 
from Eq. (17) by using the values of the model parameters listed in Table
1. Figure 5 is the same as Fig. 4 but the empty circles 
are surrogate data obtained by assuming a Student's {\it t} statistics for the 
noise terms. 

The agreement between the results obtained for the Gaussian surrogate data 
and the theoretical prediction is very 
good. This is due to the fact that the statistical uncertainty 
associated to the curve described
by Eq. (17) for a set of 1071 stocks is of the order of
the standard deviation of the random 
variable $<\epsilon_i^2(t)>$. This quantity is the square root
of Eq. (12) and it is equal to 
$3.8~10^{-5}$ for the model parameters of Table 1. This error is very small
with respect to the scale of Fig. 4 and therefore the empty circles cluster
very close to the dashed line. On the other hand for the single-index model 
with Student's {\it t} noise terms the standard deviation of 
$<\epsilon_i^2(t)>$ diverges and for this reason the dashed curve in 
Fig. 5 describes well the average behavior of the square variety 
but large fluctuations around the average behavior are observed.

We finally compare the theoretical prediction and the behavior 
of surrogate data with the results obtained in the empirical analysis 
(filled circles in Figures 4 and 5). The empirical data
are on average approximately described by the single-index model but
the fluctuation  
of the empirical results around the theoretical line is much larger 
than the one predicted by the single-index model with
Gaussian noise terms and it is larger than the one observed in the
simulated surrogate data with Student's
{\it t} noise terms. Moreover, in the presence of large values of
$|\mu(t)|$  the 
discrepancy between the empirical data and the prediction of the 
single-index model becomes progressively more pronounced. 
In particular, for large values of $|\mu(t)|$ the square 
variety is much larger than the value predicted by the single-index model.
For large values of $|\mu(t)|$, we observe a different behavior in
crash and rally days. Specifically, the variety
of the portfolio has a relative increase with respect to the theoretical
prediction of the single-index model which is larger for rally 
than for crash days.

\section{Conclusions}

The present study investigates the behavior of the variety
of a portfolio of stocks in normal and extreme days of market 
activity. The variety of a portfolio of stocks carries information
about the market activity which is not included in simple models
of financial markets such as the single-index model. 
The time evolution of the variety shows a breaking of temporal 
symmetry at the crash days. In fact aftershock periods of the 
variety are clearly observed only in empirical data whereas the 
surrogate data of the single-index model show a time evolution 
which is approximately time reversal. This is observed even if
one uses the empirical time series of the Standard and Poor's 500
index as market factor. In other words  
the time series of the variety is showing the time arrow much better
than any market factor. Independent evidence of absence of time 
reversal of the statistical properties of financial time series 
has been given in Ref. \cite{Arneodo98}

A second point considered in our study concerns the value of
the variety observed in empirical data and the difference between 
it and the value predicted by the single-index model. The single-index
model is able to mimic the average behavior of the square variety 
but fails in describing quantitatively the correct relation between the
square variety and the mean return of the ensemble distribution.
In particular the difference between empirical data and 
theoretical description is more pronounced for large positive values
of the mean return of the ensemble distribution. Other significant deviations
are also observed for extreme negative values of the mean return.
A large spreading around the theoretical curve is observed in the entire 
$\mu(t)$ axis. This spreading cannot be simply explained as statistical 
uncertainty due to the presence of noise terms. In fact surrogate data 
both with Gaussian and Student's $t$ noise terms are able to explain
only part of the spreading.

A possible interpretation of the deviations of the empirical values of
the square variety from the theoretical predictions of the single-index
model observed in crash and rally days of financial markets  
is that the portfolio of stocks undergoes a change of the 
$\beta_i$ parameters in the presence of large movements of the market.
Under this interpretation, the different behavior in crash and rally
days could reflect the existence of a different
degree of homogeneity of the market. In particular, within this framework,
empirical data indicates that the market 
inhomogeneity increases more during rally than during crash days.

\section{Acknowledgements}
The authors thank INFM and MURST for financial support. This work 
is part of the FRA-INFM project {\it Volatility in financial markets}. 
F. Lillo acknowledges FSE-INFM for his fellowships.
We wish to thank Giovanni Bonanno for help in numerical calculations.

%\newpage

\begin{figure}[t]
\epsfxsize=3in
\epsfbox{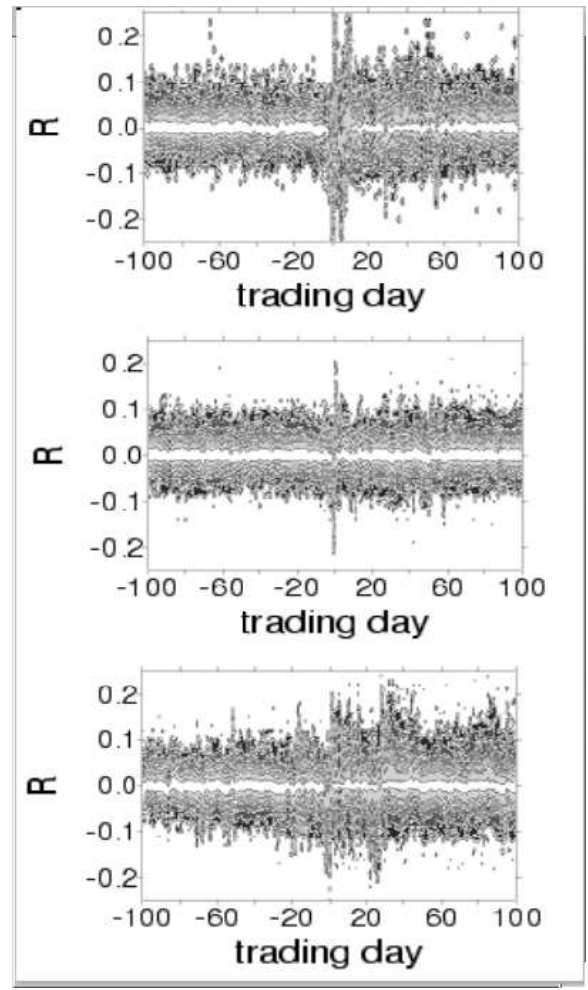}
\vspace{0.3cm}
\caption{Contour plots of the logarithm of the ensemble return distribution
in a $200$ trading days time interval centered at 19 October 1987 (top panel),
27 October 1997 (middle panel), and 31 August 1998 (bottom panel). In all the 
three panels we set the value 0 in the abscissa at the crash day. The contour 
plots are obtained for equidistant intervals of the logarithmic probability 
density. The brightest area of the contour plots corresponds to the most 
probable value.}
\label{fig1}
\end{figure} 

\begin{figure}[t]
\epsfxsize=3in
\epsfbox{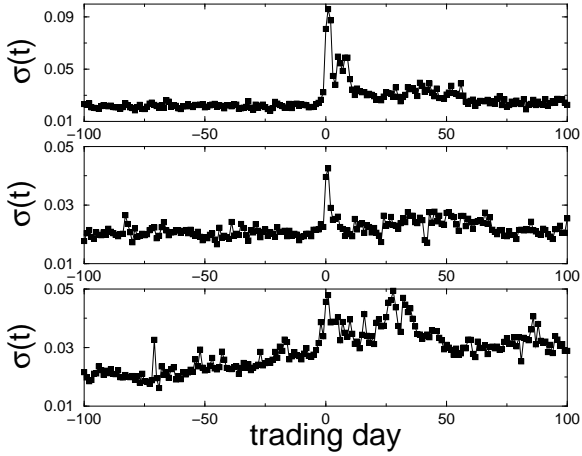}
\vspace{0.3cm}
\caption{Time series of the variety $\sigma(t)$ of the ensemble return 
distribution
in a $200$ trading days time interval centered at 19 October 1987 (top panel),
27 October 1997 (middle panel), and 31 August 1998 (bottom panel). In all the 
three panels we set the value 0 in the abscissa at the crash day.
It should be noted that the scale of the $y$-axis is twice larger for
the 1987 crisis.}
\label{fig2}
\end{figure} 

\begin{figure}[t]
\epsfxsize=3in
\epsfbox{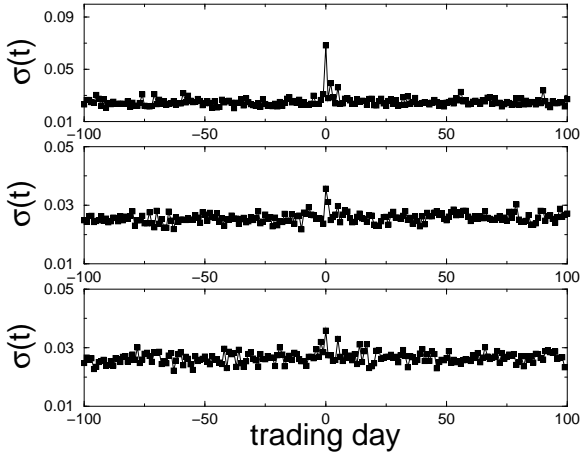}
\vspace{0.3cm}
\caption{Time series of the variety $\sigma(t)$ of surrogate data
generated by using a single-index model with Student's $t$ noise
terms with $\kappa=3$. The parameters of the single-index model
are given in Table 1 and the market factor is the Standard and Poor's 500
index. The considered time periods are the same as in Fig. 2.
They consist
in a $200$ trading days time interval centered at 19 October 1987 (top panel),
27 October 1997 (middle panel), and 31 August 1998 (bottom panel). In all the 
three panels we set the value 0 in the abscissa at the crash day.
The aftershock periods observed in empirical data are not present in
surrogate data.}
\label{fig3}
\end{figure} 

 \begin{figure}[t]
\epsfxsize=3in
\epsfbox{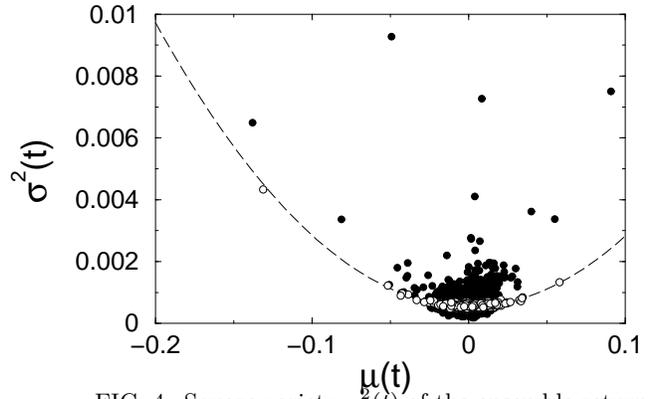}
\vspace{0.3cm}
\caption{Square variety $\sigma^2(t)$ of the ensemble return 
distribution as a function of the mean $\mu(t)$ for each trading day 
of the investigated time period. Each black circle refers to 
one trading day for empirical data. The white circles are the results
obtained by analyzing surrogate data generated according to the 
single-index model with Gaussian noise terms. The dashed line 
is the theoretical prediction of Eq. (17) with the parameters of Table 1.}
\label{fig4}
\end{figure} 

\begin{figure}[t]
\epsfxsize=3.2in
\epsfbox{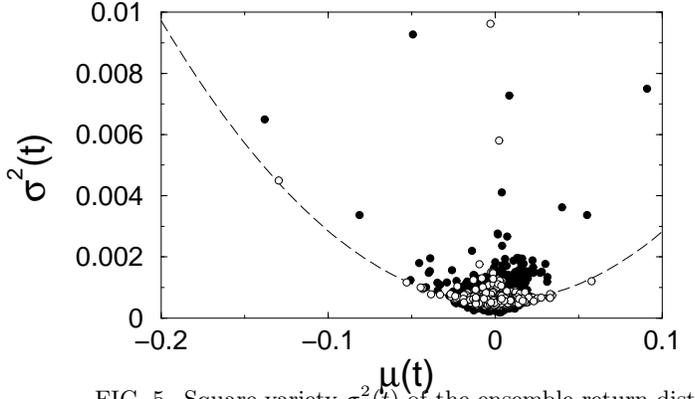}
\vspace{0.3cm}
\caption{Square variety $\sigma^2(t)$ of the ensemble return 
distribution as a function of the mean $\mu(t)$ for each trading day 
of the investigated time period. Each black circle refers to 
one trading day for empirical data. The white circles are the results
obtained by analyzing surrogate data generated according to the 
single-index model with Student's $t$ noise terms with $\kappa=3$. 
The dashed line is the theoretical prediction of Eq. (17) with 
the parameters of Table 1.}
\label{fig5}
\end{figure} 

\begin{table}
\caption{Value of the ensemble mean and standard deviation
of the single-index model parameters obtained from empirical
data with least square method by using the Standard and Poor's
500 index as market factor. The portfolio is composed by the 
1071 stocks continuously
traded in the New York Stock Exchange during the period 1987-1998.}
\label{tab:1}       % Give a unique label
% For LaTeX tables use
\begin{tabular}{ccc}
\hline\noalign{\smallskip}
Parameter &mean&standard deviation \\
\noalign{\smallskip}\hline\noalign{\smallskip}
$\alpha_i$&$2.02~10^{-4}$&$3.93~10^{-4}$\\
$\beta_i$&$6.39~10^{-1}$&$3.06~10^{-1}$\\
$\sigma^2_{\epsilon_i}$&$5.47~10^{-4}$&$6.69~10^{-4}$\\
\noalign{\smallskip}\hline
\end{tabular}
%Or use
%\vspace*{5cm}  % with the correct table height
\end{table}

\begin{table}
\caption{Time average and standard deviation of $\mu(t)$ for
empirical data, the theoretical prediction of the single-index model
(Eqs (6) and (7)) and surrogate data generated according to Eq. (4)
with Gaussian and Student's $t$ noise terms with $\kappa=3$.}
\label{tab:2}       % Give a unique label
% For LaTeX tables use
\begin{tabular}{ccc}
\hline\noalign{\smallskip}
$\mu(t)$&mean&standard deviation \\
\noalign{\smallskip}\hline\noalign{\smallskip}
data&$5.6~10^{-4}$&$73.7~10^{-4}$\\
theory&$5.7~10^{-4}$&$65.8~10^{-4}$\\
Gaussian&$5.8~10^{-4}$&$65.7~10^{-4}$\\
Student&$5.6~10^{-4}$&$65.9~10^{-4}$\\
\noalign{\smallskip}\hline
\end{tabular}
% Or use
%\vspace*{5cm}  % with the correct table height
\end{table}

\begin{table}
\caption{Time average and standard deviation of $\sigma^2(t)$ for
empirical data, the theoretical prediction of the single-index model
(Eqs (14) and (15)) and surrogate data generated according to Eq. (4)
with Gaussian and Student's $t$ noise terms with $\kappa=3$.}
\label{tab:3}       % Give a unique label
% For LaTeX tables use
\begin{tabular}{ccc}
\hline\noalign{\smallskip}
$\sigma^2(t)$&mean&standard deviation \\
\noalign{\smallskip}\hline\noalign{\smallskip}
data&$5.4~10^{-4}$&$3.8~10^{-4}$\\
theory&$5.8~10^{-4}$&$8.5~10^{-5}$\\
Gaussian&$5.6~10^{-4}$&$8.4~10^{-5}$\\
Student&$5.6~10^{-4}$&$2.2~10^{-4}$\\
\noalign{\smallskip}\hline
\end{tabular}
% Or use
\vspace*{5cm}  % with the correct table height
\end{table}

\end{document}